\documentclass{appolb}
\usepackage{graphicx}
\usepackage{authblk}
\usepackage{verbatim}
\usepackage{fancyhdr}

\begin{document}

\title{Cosmic Ray Extremely Distributed Observatory: a global network of detectors to probe contemporary physics mysteries%
\thanks{Presented at “Excited QCD 2018”, Kopaonik, Serbia, March 11-15, 2018.}
}


\author[1]{K. Almeida Cheminant\thanks{corresponding author: kevin.almeida-cheminant@ifj.edu.pl}}
\author[2]{\L{}. Bratek}
\author[3]{D. E. Alvarez-Castillo}
\author[1]{N. Dhital}
\author[1]{D. G\'{o}ra}
\author[1]{P. Homola}
\author[1]{R. Kami\'{n}ski}
\author[4]{M. Kasztelan}
\author[1]{K. Kopa\'{n}ski}
\author[5]{P. Kovacs}
\author[1]{M. Krupi\'{n}ski}
\author[6]{M. Magry\'{s}}
\author[7]{M. Marek}
\author[1,3]{V. Nazari}
\author[2]{M. Nied\'{z}wiecki}
\author[1,2,8]{W. Noga}
\author[6]{K. Oziomek}
\author[6]{M. Pawlik}
\author[2]{K. Rzecki}
\author[9]{J. Zamora-Saa}
\author[10]{F. Simkovic}
\author[1,2]{K. Smelcerz}
\author[11]{K. Smolek}
\author[1]{J. Staliesak}
\author[1]{O. Sushchov}
\author[1]{K. Wo\'{z}niak}
\affil[1]{Institute of Nuclear Physics Polish Academy of Sciences, Radzikowskiego 152, Cracow, Poland}
\affil[2]{Cracow University of Technology, Faculty of Physics, Mathematics and Computer Science, Warszawska 24, Cracow, Poland}
\affil[3]{Joint Institute for Nuclear Research, Dubna 141980, Russia}
\affil[4]{National Centre for Nuclear Research, Andrzeja So\l{}tana 7, 05-400 Otwock-\'{S}wierk, Poland}
\affil[5]{Institute for Particle and Nuclear Physics,Wigner Research Centre for Physics, Hungarian Academy of Sciences, H-1525 Budapest, Hungary}
\affil[6]{ACC Cyfronet AGH-UST, Nawojki 11, 30-950 Cracow, Poland}
\affil[7]{Amateur astrophysicist}
\affil[8]{Stowarzyszenie POLARIS - OPP, Sopotnia Wielka 174, Jeleśnia, Poland}
\affil[9]{Departamento de Ciencias Físicas, Universidad Andres Bello, Sazié 2212, Piso 7, Santiago, Chile}
\affil[10]{Comenius University, Faculty of Mathematics, Physics and Informatics, Mlynska dolina, SK-842 48 Bratislava, Slovakia}
\affil[11]{Institute of Experimental and Applied Physics, Czech Technical University in Prague, Horská 3a/22, 128 00 Prague 2, Czech Republic}
\renewcommand\Authands{ and }

\maketitle
\newpage
\begin{abstract}
In the past few years, cosmic-rays beyond the GZK cut-off ($E > 5 \times 10^{19}$ eV) have been detected by leading collaborations such as Pierre Auger Observatory. Such observations raise many questions as to how such energies can be reached and what source can possibly produce them. Although at lower energies, mechanisms such as Fermi acceleration in supernovae front shocks seem to be favored, top-down scenarios have been proposed to explain the existence of ultra-high energy cosmic-rays: the decay of super-massive long-lived particles produced in the early Universe may yield to a flux of ultra-high energy photons. Such photons might be presently generating so called super-preshowers, an extended cosmic-ray shower with a spatial distribution that can be as wide as the Earth diameter. The Cosmic Ray Extremely Distributed Observatory (CREDO) mission is to find such events by means of a network of detectors spread around the globe. CREDO's strategy is to connect existing detectors and create a worldwide network of cosmic-ray observatories. Moreover, citizen-science constitutes an important pillar of our approach. By helping our algorithms to recognize detection patterns and by using smartphones as individual cosmic-ray detectors, non-scientists can participate in scientific discoveries and help unravel some of the deepest mysteries in physics.
\end{abstract}

\PACS{DOI:10.5506/APhysPolBSupp.11.489}

\section{Introduction}
\begin{figure}[htb]
\centerline{%
\includegraphics[height=6cm]{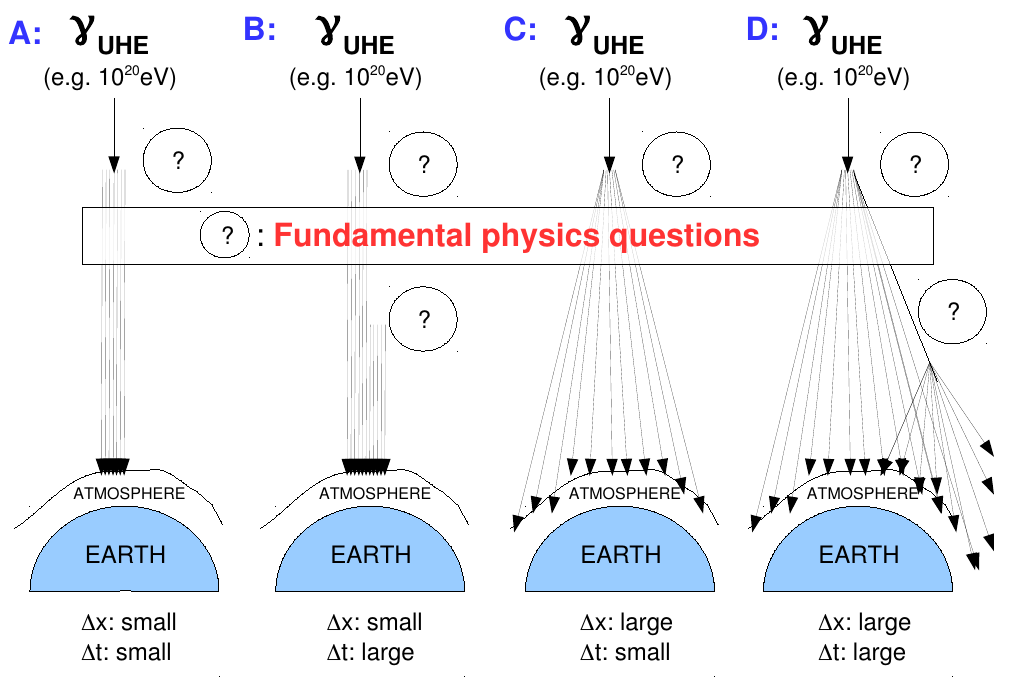}}
\caption{Classification of SPSs: different classes correspond to different spatial and temporal distribution of the particles at the top of the atmosphere \cite{dhital17,sushchov17}.}
\label{Fig:F1}
\end{figure}

Since the discovery of ionizing radiation from outer space - the so-called cosmic-rays (CRs) - by Victor Hess in the early twentieth century, a series of questions have kept contemporary astrophysicists inquisitive. Although it is strongly believed that CRs up to $10^{15}$ eV are accelerated at the front shock of supernovae through first-order Fermi acceleration \cite{bell78,tanimori98}, it has yet to be understood how particles can reach energies above $10^{18}$ eV - named ultra-high energy cosmic-rays (UHECRs) - as observed by the Pierre Auger Collaboration \cite{auger10-1}. A few sources such as active galactic nuclei and gamma-ray bursts are among the few candidates capable of accelerating cosmic-rays at these energies through bottom-up mechanism (see review in \cite{bhattacharjee00}). However, latest observations show a weak correlation between arrival directions of UHECRs and location of known mentioned sources \cite{auger10-2,abbasi14}. As a consequence, it is interesting to consider other potential origins such as top-down scenarios. In these models, a supermassive particle produced in the early Universe decays to UHECRs and ultra-high energy photons (UHE-$\gamma$)\cite{berezinsky97,auger17}. The existence of UHE-$\gamma$ does not solely arises from particle decays but also from other mechanisms such as GZK effect \cite{greisen66,zatsepin66} where a UHE proton loses energy by interacting with the cosmic microwave background, leading to the production of UHE-$\gamma$. Considering these different processes, it seems realistic to assume the existence of a UHE-$\gamma$ flux on Earth. Latest report from the Pierre Auger collaboration \cite{auger17} gives upper limits on the photon flux that put severe constraints on top-down scenarios (Figure 6 in \cite{auger17}).
\\
One could conclude that extinction of the photon flux is important enough to prevent any UHE-$\gamma$ to reach the Earth. In fact, a UHE-$\gamma$ propagating through the Universe may interact with radiation fields and generate an electromagnetic cascades of secondary particles called \textit{super-preshower} (SPS). In this turn of events, a collection of space and time-correlated particles arrive at the top of the Earth's atmosphere. Because spatial distribution of SPS particles can be as wide as the Earth's diameter (see Figure \ref{Fig:F1} taken from \cite{dhital17,sushchov17}), a global network of connected detectors looking for correlated air showers seems necessary. Inaugurated in August 2016, the Cosmic-Ray Extremely Distributed Observatory (CREDO) \cite{dhital17,sushchov17} focuses its efforts on detecting such cosmic-ray ensembles. However, building a new worldwide infrastructure dedicated to this task would be both time consuming and particularly costly. For this reason, CREDO's strategy relies on two key aspects: using data from existing observatories such as gamma-ray telescopes, neutrino detectors and cosmic-ray arrays, and the conception of cheap cosmic-rays detectors easy to manufacture. The enormous amount of data expected to be obtained gives the opportunity to open the project not only to the whole scientific community but also to the general public. Through diverse projects, from data acquisition to data analysis, citizen science is put center stage as a direct outcome of outreach and public engagement.

\section{Simulations: SPSs scenarios and imprints}
\begin{figure}[htb]
\centerline{%
\includegraphics[width=12.5cm]{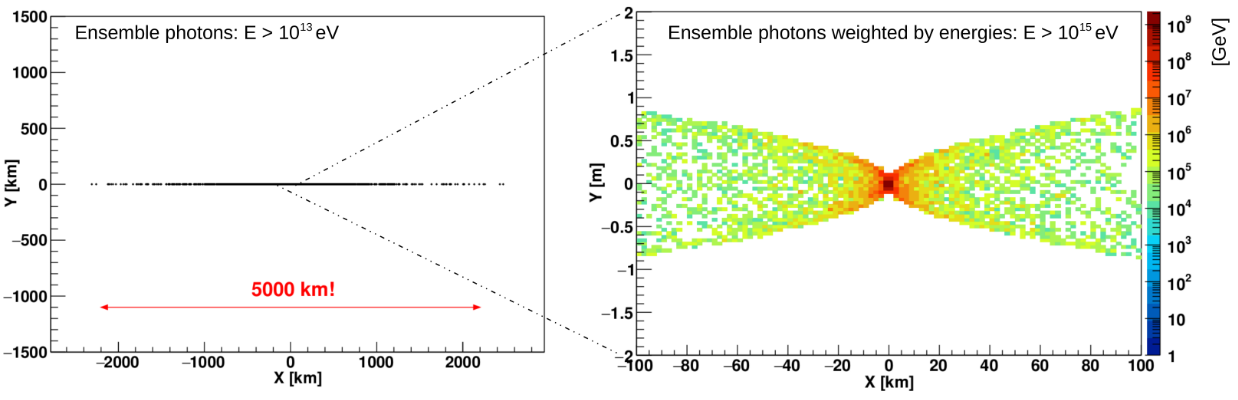}}
\caption{A UHE-$\gamma$ can interact with the Sun's magnetic field and produce a class C (see Figure \ref{Fig:F1}) super-preshower. These plots show the distribution of particles at the top of the Earth's atmosphere for particles with energy $E > 10^{13}$ eV (left) and $E > 10^{15}$ eV (right).}
\label{Fig:F2}
\end{figure}

The pioneer research direction taken by the CREDO collaboration inevitably implies a lack of knowledge regarding the characteristics of SPSs and therefore, the necessity to undertake a theoretical work in order to understand how SPSs can be formed, what are their signatures on Earth and how the global network set up by CREDO can detect such events. For this purpose, several projects are currently being conducted.
\\
When a UHE-$\gamma$ interacts with a local magnetic field, it can produce a pair of electron-positron which radiates bremsstrahlung photons as they are deviated from the original UHE-$\gamma$ propagation direction by the magnetic field. This effect is known as the preshower effect \cite{mcbreen81} and has been simulated in the case of a UHE-$\gamma$ interacting with the Sun's magnetic field. Figure \ref{Fig:F2}  shows the energy distribution of particles above $10^{13}$ eV obtained with the PRESHOWER algorithm \cite{homola05} in this scenario. Bremsstrahlung photons of the highest energies are mostly localized near the core of the shower and the East-West distribution is elongated on several thousands of kilometers. This is the chance for a very unique signature: one can notice the wide energy range, from TeV to EeV, implying that any structure capable of detecting air showers particles such as gamma-ray telescopes and cosmic-ray detectors is relevant to CREDO's ambition.
\\

Another ongoing research project is focusing on the possibility to detect class A (see Figure \ref{Fig:F1}) SPSs by the means of gamma-ray telescopes \cite{almeida17}, with a particular attention put on the preshower effect obtained from the geomagnetic field. In this study, the air showers generated by the particles at the top of the atmosphere produce Cherenkov radiation that is detected by the telescopes camera. The images formed on the cameras being dependent of the primary particle, the signature of SPSs could possibly be identified if it can be discriminated from the cosmic-ray background. Such discrimination could be obtained by analyzing the images parameters - so-called Hillas parameters \cite{hillas85} - through a multivariate analysis that is currently being worked out. 
\thispagestyle{empty}

\section{Citizen science and data analysis}
In order to reach the goals set by CREDO, having a global network covering a large geographical area with as many detectors as possible becomes one of the pillar of the overall strategy. The resulting stream of data opens opportunities beyond the scientific community: the general public gets a chance at being directly involved in scientific discoveries, not only as simple participants but also as co-authors.  The scale of commitment offered by CREDO ranges from data collection with simple devices to data analysis through platforms already known and used worldwide.
\\
To begin with, CREDO has launched its own mobile application which converts the camera of a smartphone into a particle detector. Although similar applications like DECO \cite{meehan17} and CRAYFIS \cite{whiteson14} already exist on the market, CREDO's version differs by its open-source nature, therefore encouraging users to participate in its improvement. In order to make this experiment more than a simple passive participation, interactive tools are available for the user such as keeping track of his detections and seeing his ranking among other users. After passing through some filters removing possible artifacts, the data is periodically transfered on a server at ACC Cyfronet AGH-UST where data from other participating detector arrays are also stored. 
\begin{figure}[t]
\centerline{%
\includegraphics[width=12.5cm]{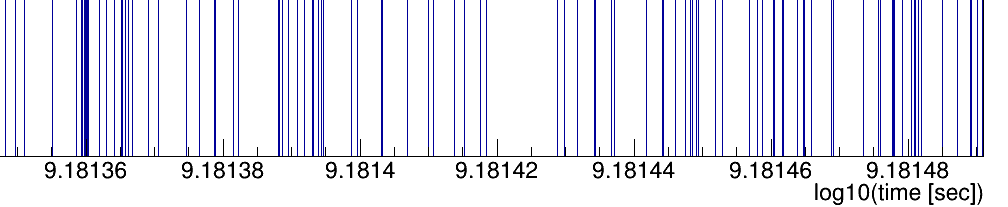}}
\caption{Distribution of 100 detections timestamps from smartphone user corresponding to about 6 days of taking data with the CREDO app. Some excess seems to appear at the beginning of the data acquisition. Time is given in UNIX format.}
\label{Fig:F3}
\end{figure}
\\
Several experiments such as looking for excesses in the number of detections (see Figure \ref{Fig:F3}) on a fixed time window are being proposed to the general public. In this study, one is looking for flares of events based only on the timestamps of the detections as a first approximation. If such excess is significantly above the random background distribution, more sophisticated analysis taking into account localization can be performed. The results give the opportunity to give a direct feedback to the application users, both on an individual and global scale.
\\
From the previous analysis, the most significant flares can be used to generate maps (see Figure 4 of \cite{homola17}). These maps can be classified by human eyes in the Dark Universe Welcome project of the Zooniverse citizen science platform in order to identify strange patterns that could be peculiar features of SPSs. In this endeavour, the manpower offered by non-scientists is critical: classifying patterns that remain unknown to physicists and that could be the sign of new physics.

\section{Conclusion}
By establishing a worldwide infrastructure of CRs detectors, CREDO aims at opening a new channel of observation on the Universe: the channel of cosmic-ray ensembles. Instead of looking for single particles at the top of the atmosphere, one searches for several correlated particles that might be of a common origin. Achieving such ambitious goals requires efforts both on theoretical and experimental levels and therefore, calls for the participation of both scientific community from all horizons and the general public. Studying the signature left by SPSs on detectors is done along with developing tools for both data collection and analysis. Non-scientists are invited to actively contribute to both the latter stages by using their smartphones as cosmic-ray detectors and classifying the obtained data on an online platform. The discovery potential offered by the CREDO project provides a fertile ground for educational purposes and collaborations beyond the field of cosmic-ray research. 
\\
\\
This research has been supported in part by PLGrid Infrastructure. We warmly thank the staff at ACC Cyfronet AGH-UST for their always helpful supercomputing support. The CREDO application is developed in Cracow University of Technology. The Dark Universe Welcome citizen science experiment was developed with the help of the ASTERICS Horizon2020 project. ASTERICS is a project supported by the European Commission Framework Program Horizon 2020 Research and Innovation action under grant agreement n.653477.

\end{document}